\documentclass{emulateapj}
\usepackage{alltt}
\journalinfo{ApJ}

\shorttitle{Comment on arXiv}
\shortauthors{Bulbul et al}

\begin{document}

\title{Comment on ``Dark matter searches going bananas: the contribution of
  Potassium (and Chlorine) to the 3.5 keV line''}

\author{Esra Bulbul (1), Maxim Markevitch (2), Adam R. Foster (1), Randall K.
  Smith (1), Michael Loewenstein (2), Scott W. Randall (1)}
\affil{(1) Harvard-Smithsonian Center for Astrophysics, (2) NASA/GSFC}

\begin{abstract}
The recent paper by \citet{Jeltema2014} claims that contributions from
\ion{K}{18} and \ion{Cl}{17} lines can explain the unidentified emission
line found by \citet{Bulbul2014} and also by
\citet{Boyarsky2014a,Boyarsky2014b}. We show that their analysis relies upon
incorrect atomic data and inconsistent spectroscopic modeling. We address
these points and summarize in the appendix the correct values for the
relevant atomic data from AtomDB.
\end{abstract}

\section{Introduction}

In a recent preprint ``Dark matter searches going bananas: the contribution
of Potassium (and Chlorine) to the 3.5 keV line,'' \citet[hereafter
JP]{Jeltema2014} claim that the unidentified $E\approx 3.55-3.57$ keV
emission line that we detected in the stacked galaxy cluster spectra
\citep[hereafter B14]{Bulbul2014} and \citet{Boyarsky2014a} detected in
Perseus and M31 (as well as their more recent detection of the same line in
the Galactic Center, \citep{Boyarsky2014b}) can be accounted for by an
additional \ion{Cl}{17} Ly$\beta$\ line and by broadening the model
uncertainty for the flux of the \ion{K}{18} He-like triplet. These
transitions occur at $E\approx 3.51$ keV, close to our unidentified line.
In B14, we considered the K line among other possibilities and concluded
that it cannot explain the new line. Here we respond to JP's concerns,
focusing on our galaxy cluster analysis.

Specifically, JP raise three key points about the analysis in B14:
\begin{enumerate}
\item A possible \ion{Cl}{17} Ly$\beta$\ line at $E=3.51$ keV was not
included in our model;
\item The plasma temperatures derived from the ratios of fluxes of
\ion{S}{16}, \ion{Ca}{19} and \ion{Ca}{20} lines in the cluster spectra are
inconsistent, thus a much larger range of temperatures must be allowed in
modeling;
\item When using a wider range of possible temperatures, and scaling from
the fluxes for the \ion{S}{16}, \ion{Ca}{19}, \ion{Ca}{20} lines reported by
B14 for the Perseus cluster, the total flux in the \ion{K}{18} and
\ion{Cl}{17} lines can match that of the unidentified line.
\end{enumerate}
They conclude that, accounting for these points, no additional line is
required by the B14 data. We address these items below.

\subsection{Atomic Data}
\label{sec:emissivities}

In a study of this nature, using accurate atomic data is vital. JP state
that they have used AtomDB \citep{Smith2001} to calculated their line
fluxes. Though they do not cite the version, from the fact that they used
the recently added lines of Chlorine, it must be the latest version 2.0.2
\citep{Foster2012}. However, we have been unable to recreate the line ratios
in Table 3 of JP using AtomDB v2.0.2. In theory, these should be the fluxes
from their Table 2, multiplied by the ratio of predicted \ion{K}{18}
emissivities to that of the line in question.

We can, however, recreate their Table 3 if we use the approximate values
available in the ``strong lines'' option at
http://www.atomdb.org/WebGUIDE/webguide.php. As described on that page, this
option uses an approximation
\begin{equation}{
\epsilon (T) = \epsilon (T_{peak})  N(T)/N(T_{peak})}
\end{equation}
where $\epsilon$ is the emissivity, $T$ is the requested temperature,
$T_{peak}$ is the temperature for which the transition's emissivity is its
maximum, and $N$ is the abundance of the ion. This approximation is intended
for quick identification of possible strong lines, as it disregards the change in line emissivity with temperature, instead accounting only for the relative change in ion abundance.%
\footnote{A note that accompanies the results of every line search on that
  web page further states: ``The emissivities listed here are intended only
  as a guide, and should not be used for analysis ... For correct
  emissivities, please use the full AtomDB database.}

\begin{figure}
\begin{center}
\includegraphics[width=0.85\columnwidth]{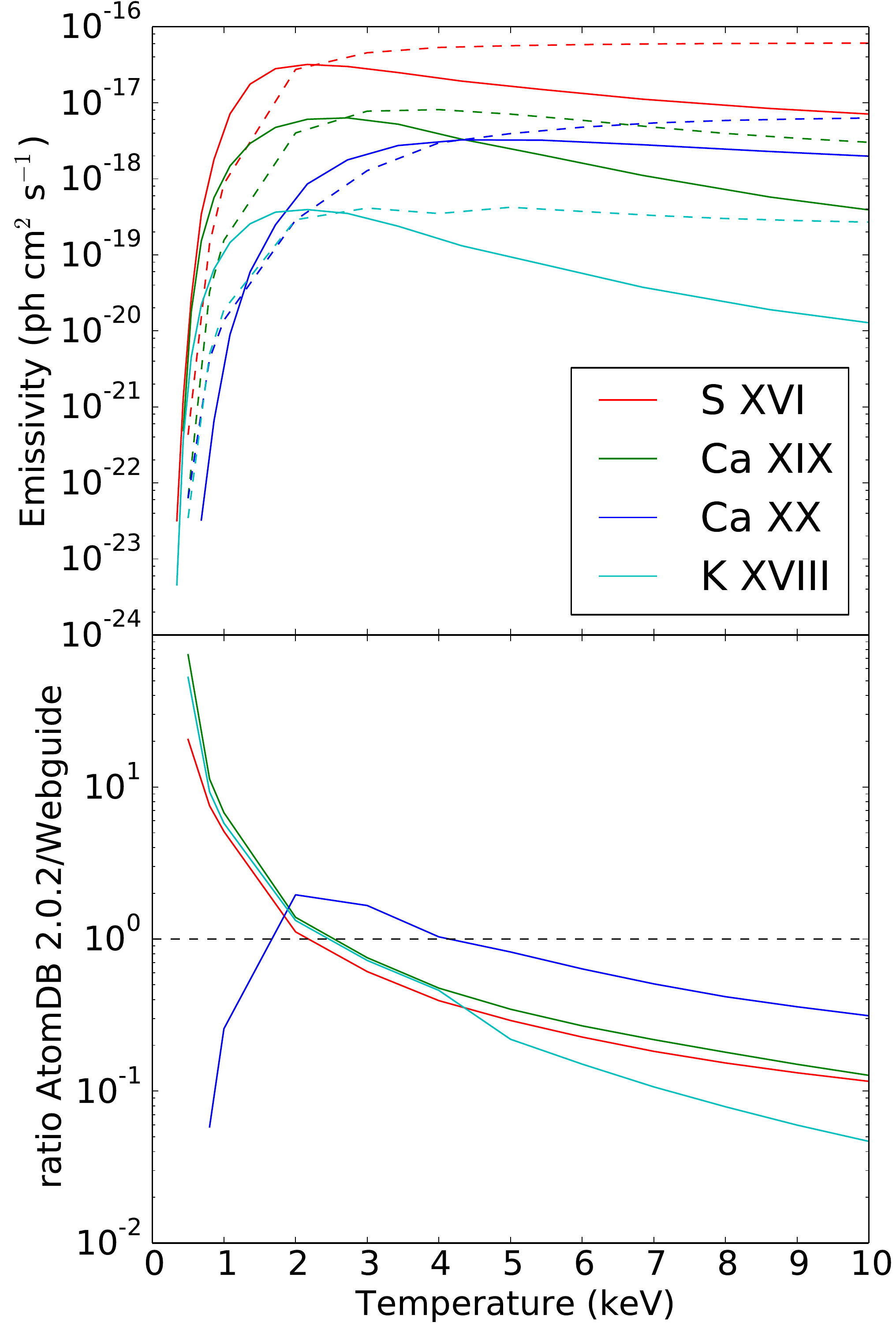} 

\caption{Upper panel: emissivities for the relevant lines as a function of
  temperature. Solid curves show the correct data from AtomDB 2.0.2, and
  dashed lines show those from the WebGUIDE ``strong lines'' approximation.
  Lower panel: ratios of the correct emissivities to the ``strong line''
  emissivities for the same lines.}

\label{fig:webguide}
\end{center}
\end{figure}

Using these approximate data, we were able to recreate the values in JP's
Table 3 exactly from the data in Table 2, to identify exactly which lines JP
included in their flux ratio calculations, and to explain the line ratios
discussed in their \S 3.1. The error due to the use of this approximation
can be very large for temperatures away from the line peak emissivity
temperature, as illustrated in Fig.\ \ref{fig:webguide} for our four
relevant lines.

\subsection{Line Ratios as Temperature Diagnostics}
\label{sec:ratios}

\begin{figure}
\begin{center}
\includegraphics[width=0.85\columnwidth]%
{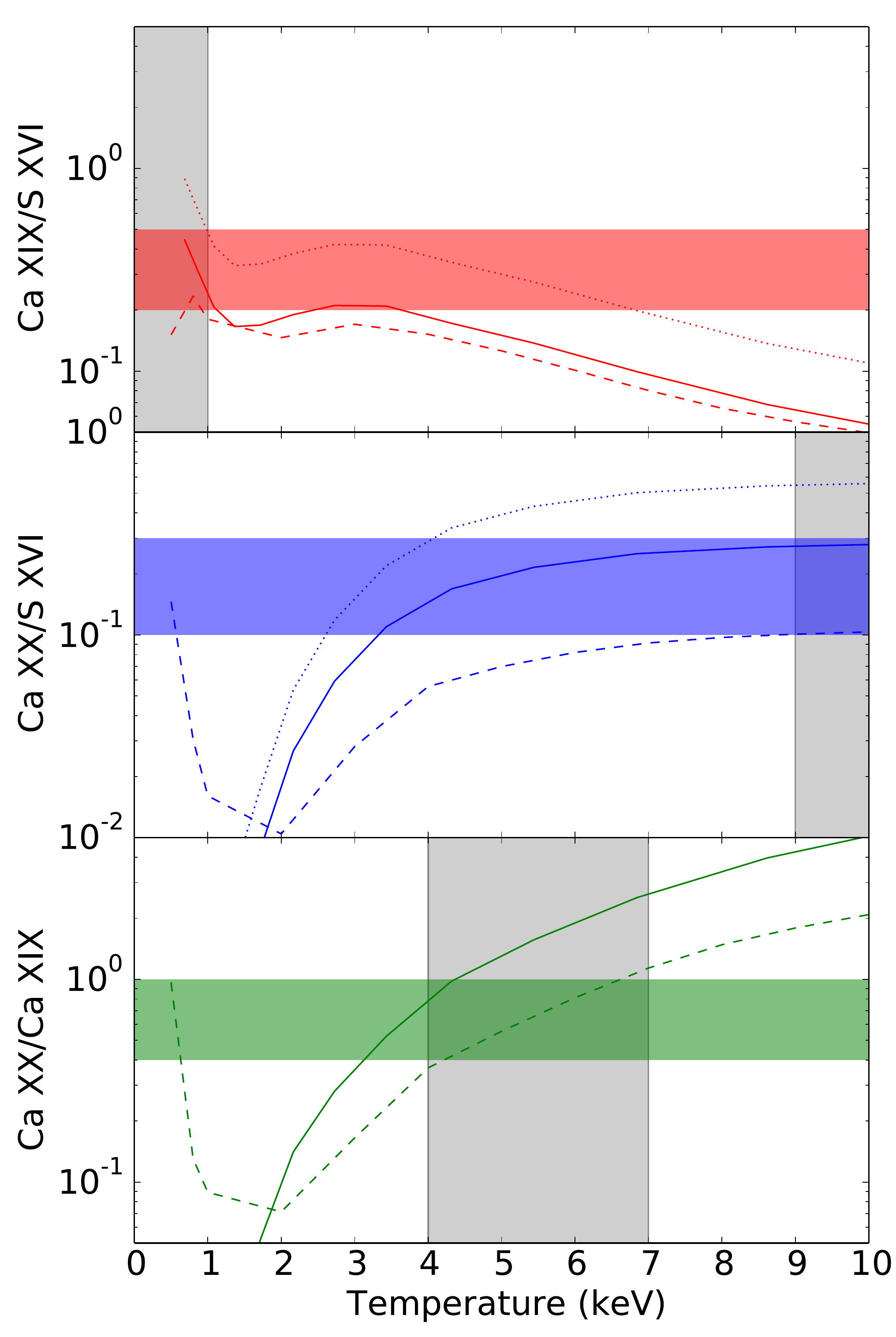} 

\caption{The line ratios for \ion{S}{16}, \ion{Ca}{19}, \ion{Ca}{20} from
  (solid line) AtomDB 2.0.2, (dotted line) AtomDB 2.0.2, assuming a sulfur
  abundance of 0.5 solar, and (dashed line) from the strong line
  approximation, as used by JP. The horizontal shaded bars show the range of
  observed values for the various cluster subsamples in B14.
  The gray shaded intervals show the valid temperature ranges as given by
  JP.}

\label{fig:ratios}
\end{center}
\end{figure}

\begin{figure}
\begin{center}
\includegraphics[width=1.0\columnwidth]{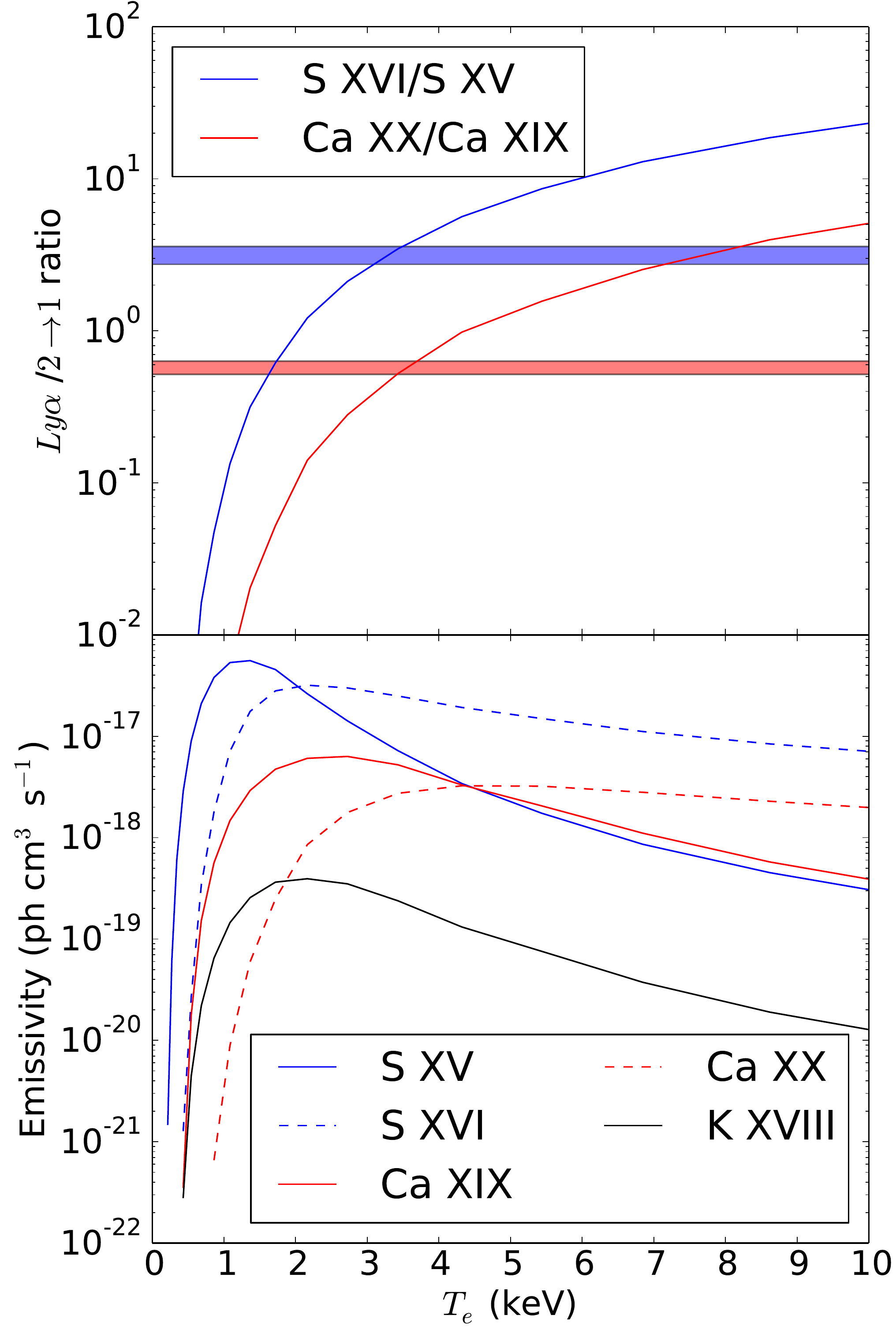} 

\caption{\textit{Top}: Emissivity ratios for \ion{S}{16}/\ion{S}{15} and
  \ion{Ca}{20}/\ion{Ca}{19}, with the observed values from the Perseus MOS
  full-cluster data shown by horizontal bands (representing 90\% intervals).
  Both observed line ratios indicate the same $T\approx 3.5$ keV.
  \textit{Bottom}: The emissivities of the above four features and the
  \ion{K}{18} line vs.\ temperature, from AtomDB 2.0.2.}

\label{fig:s_ca_ratios}
\end{center}
\end{figure}

Incorrect atomic data easily lead to incorrect conclusions about the gas
temperature structure based on the observed line ratios. In particular, JP
find that the observed ratios of the \ion{S}{16}, \ion{Ca}{19}, \ion{Ca}{20}
lines (the lines used in B14 to estimate the \ion{K}{18} flux) indicate very
different plasma temperatures. (Of course, in a single-component plasma in
ionization equilibrium, all line ratios must correspond to the same
temperature.) Therefore, they conclude that the plasma has to have a very
complex temperature structure, and so B14 were not justified to restrict the
temperature range for our estimates of the \ion{K}{18} flux.  We will
address the K line in the next section, and here we check if the relevant
line ratios are indeed in disagreement.

Figure \ref{fig:ratios} shows the line ratios of the above 3 lines as a
function of temperature, assuming solar photospheric abundances
\citep{AG89}, using the correct AtomDB data (solid curves, our calculation)
and the WebGUIDE approximation used by JP (dashed curves).  Colored
horizontal bands show the ranges of the observed ratios for the various
cluster subsamples given in B14. Vertical gray bands show JP's temperature
ranges implied by these observed ratios (based on the intersection of the
dashed theoretical curves with the observed bands), which are indeed very
different.  However, the correct line ratios aren't quite as inconsistent
with each other; in fact, with a reasonable (factor $\sim 2$) reduction of
the relative S/Ca abundance (dotted curves in the two upper panels), all
three agree with the observed range of ratios around $T\sim 3-4$ keV.

However, to exclude the effects of relative elemental abundances on line
diagnostics of the plasma temperature, it is best to use the line ratios for
different ions of {\em the same}\/ element. Since we (and JP) are most
concerned with the presence of cool plasma components, a particularly useful
diagnostic is the \ion{S}{15} $n=2\rightarrow 1$ triplet at $E=2.45$ keV. It
should be very strong in sub-3 keV plasma --- as shown in the lower panel of
Fig.\ \ref{fig:s_ca_ratios}, it exceeds the already strong \ion{S}{16} line
at $E=2.62$ keV once the temperature drops below 2 keV. The upper panel of
Fig.\ \ref{fig:s_ca_ratios} shows the line ratios for
\ion{S}{16}/\ion{S}{15} and \ion{Ca}{20}/\ion{Ca}{19} (the latter is the
same ratio shown in the bottom panel of Fig.\ \ref{fig:ratios}). The color
bands overplot the observed ratios for the Perseus MOS spectrum from the
whole cluster (i.e., including the cool core), which should have a
contribution from cool components. Yet, the two ratios show a remarkable
agreement at $T\approx 3.5$ keV (which happens to be one of the continuum
model components, see Table 2 of B14). This indicates that (a) the
components emitting the bulk of the S and Ca lines have the same temperature
and (b) any significant contribution from the components with $T<2.5$ keV is
excluded. Note that, while both Ca lines have very low emissivities at
$T\sim 1$ keV (lower panel in Fig.\ \ref{fig:s_ca_ratios}) and one might
argue that the \ion{Ca}{20}/\ion{Ca}{19} ratio is insensitive to the
presence of components at such low temperatures in a multi-temperature
plasma, the \ion{S}{15} line is very strong at $T\sim 1$ keV, and so the S
line ratio would be biased toward that component if it is present in the
mix. In all the subsamples analyzed in B14, the \ion{S}{16}/\ion{S}{15}
ratio is above 1.8, which similarly excludes any large contributions from
cool gas. Of course, we do know that cool-core clusters have a wide range of
temperatures --- but the relative contribution of the cool components into
the emission of the relevant lines is small.

An independent consideration is the observed absolute line fluxes. Because
the \ion{Ca}{20}, \ion{Ca}{19} and \ion{S}{16} emissivities drop steeply at
low temperatures (lower panel in Fig.\ \ref{fig:s_ca_ratios}), any cool
component would have to have a very high abundance of those elements to
contribute significantly to the observed line fluxes. For example, to
produce all of the observed \ion{Ca}{20} line in the Perseus MOS spectrum
with a $T=1$ keV plasma, the Ca abundance would have to be over 100 times
solar (which is unlikely given the observed values of $0.3-2$ solar in
clusters, including their cool cores).

As a side note, the lower panel of Fig.\ \ref{fig:s_ca_ratios} also shows
the emissivity of the \ion{K}{18} line alongside the S and Ca ions. The S
and Ca lines peak at similar temperatures and straddle the peak of the K
line, which is why these ions are a particularly good predictor for
\ion{K}{18} and were used for this purpose in B14.


\subsection{Potassium}
\label{sec:potassium}

The possible contribution of \ion{K}{18} to the spectrum was a major concern
in B14, hence the extensive discussion of the potential contribution from this
line, and its inclusion in all fits. Therefore, the relevant question is not the strength
of the \ion{K}{18} line relative to the unidentified line, but whether or
not the fits require an unidentified line {\it in addition to}\/ the
\ion{K}{18}.
 
We consider the specific case highlighted by JP, that of the \ion{K}{18}
line in the Perseus MOS observations. For all \ion{K}{18} flux estimates
here, we will use the sum of the 3.47 keV and 3.51 keV components, as in JP.
Looking again at Fig.\ \ref{fig:s_ca_ratios}, the observed
\ion{S}{16}/\ion{S}{15} and \ion{Ca}{20}/\ion{Ca}{19} ratios indicate a
remarkably consistent $T\approx 3.5$ keV. If the underlying temperature is
indeed 3.5 keV, the implied \ion{K}{18} triplet flux, assuming solar
elemental abundance ratios, is $1.05\pm 0.06\times 10^{-5}$ ph cm$^{-2}$
s$^{-1}$ if derived from Ca, and $0.74\pm 0.02\times 10^{-5}$ ph cm$^{-2}$
s$^{-1}$ if derived from S.

In Table 3 of B14, the Perseus MOS has a total predicted flux for the
\ion{K}{18} line of $0.64\pm0.34 \times 10^{-5}$\,ph\,cm$^{-2}$s$^{-1}$ (the
difference from the above values, within the uncertainty, is due to the
two-component modeling and the averaging over three lines in B14). Given the
uncertainties involved in this prediction (e.g., the relative element
abundances), we applied very broad bounds on our line fits, allowing them to
range from $0.1-3\times$\ our predicted values --- including their errors.
(Note from Fig.\ \ref{fig:s_ca_ratios} that the maximum emissivity of the
\ion{K}{18} for any temperature is less than a factor 2 above the value at
the temperature given by the above line ratios.) Thus, our fit allowed the
\ion{K}{18} to rise to $3\times10^{-5}$\,ph\,cm$^{-2}$s$^{-1}$ in this
spectrum.

With this cap on the \ion{K}{18} line, the spectrum did require the
additional line at $E\approx 3.57$ keV. In \S3.4 of B14 we performed several
tests removing the caps on the \ion{K}{18} line and the \ion{Ar}{17} DR line
at a higher energy, and concluded (see also \S6 in B14) that the new line is
not significantly detected only if both these lines are allowed to be above
their upper limits by large factors.

Note that, although JP did not comment on this, their highest predicted
\ion{K}{18} flux based on the \ion{Ca}{19} line (from the Perseus flux in
Table 2 of B14 and the ratios from Tables 3 and 2 in JP --- that is, using
the incorrect atomic data) is $3.1\times 10^{-5}$ ph\,cm$^{-2}$s$^{-1}$,
assuming a temperature of 1 keV. This flux is in the range that we allowed
for \ion{K}{18} in the fit in B14 (see above). Estimates based on the
\ion{Ca}{20} line at low temperatures are irrelevant (because of the
exorbitant Ca abundance required, see \S\ref{sec:ratios} above); at higher
temperatures, their estimates based on \ion{Ca}{20} (as well as the
estimates from \ion{S}{16} for all temperatures) were again within our
allowed bounds for \ion{K}{18}.

\subsection{Chlorine}
\label{sec:chlorine}

\begin{figure}
\begin{center}
\includegraphics[width=1.0\columnwidth]{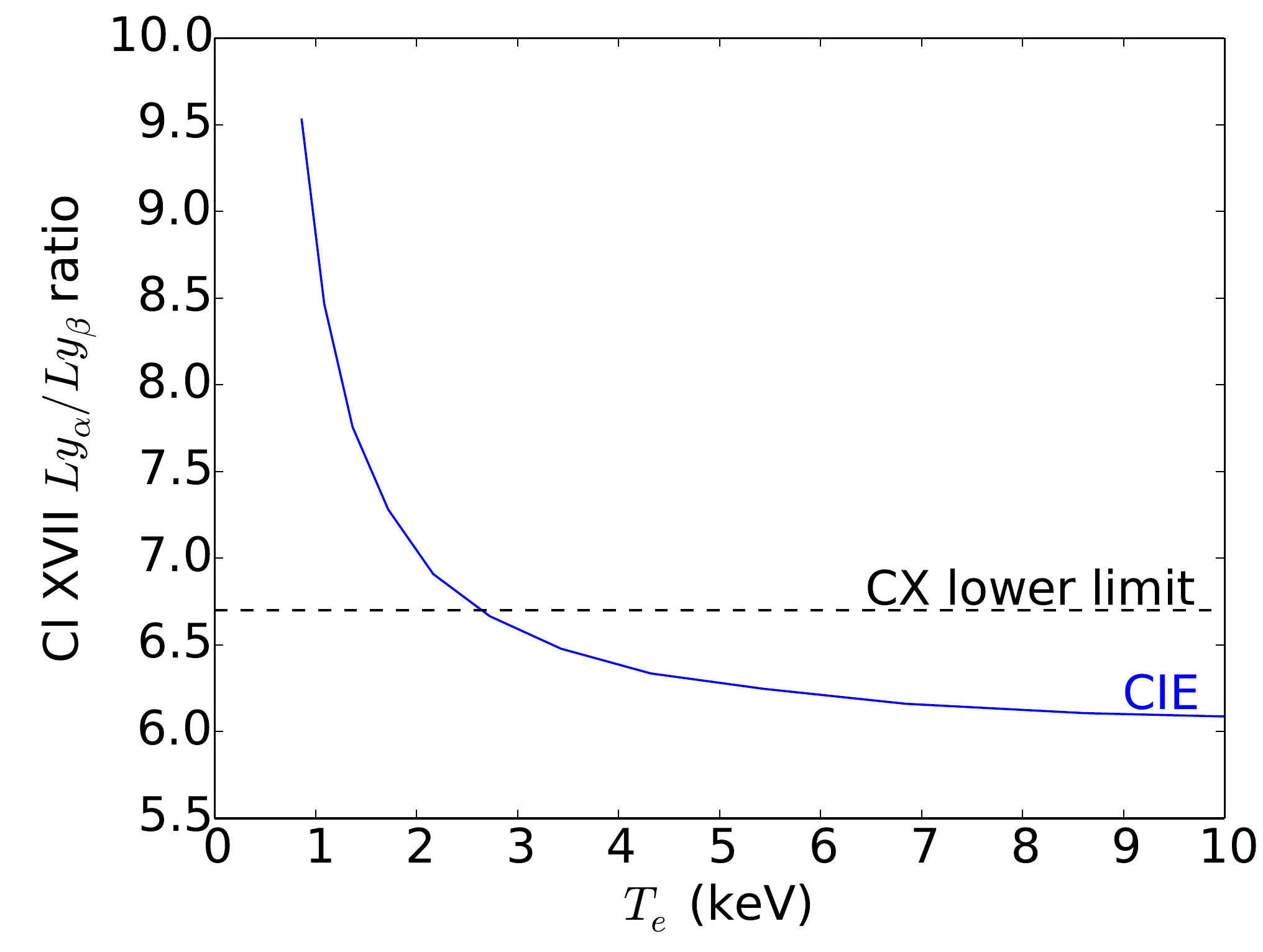} 
\caption{The \ion{Cl}{17} Ly$\alpha$/Ly$\beta$\ ratio from AtomDB 2.0.2
\label{fig:chlorine}
}
\end{center}
\end{figure}

The \ion{Cl}{17} line in question is the Ly$\beta$\ doublet. In a thermal
plasma, it is always much weaker than the Ly$\alpha$\ doublet at $E=2.96$
keV. A theoretical ratio of these lines as a function of temperature is
shown in Figure \ref{fig:chlorine} for collisional ionization equilibrium
(CIE), taken from AtomDB 2.0.2 \citep{Foster2012}. For completeness, we also
show a limit corresponding to charge exchange (CX), another possibly
relevant nonthermal mechanism (although, as discussed in B14, there has been
no evidence for any significant CX emission in clusters). The CX models of
\citet{Smith2014} do not contain Cl, so we took it to be equal to the limit
for S and Ar, which should be similar. For all temperatures, both CIE and CX
Ly$\alpha$/Ly$\beta$\ ratios are above 6.

We did not include the Cl XVIII lines in B14, because their expected fluxes
(based on the APEC CIE model) were below our threshold, but we can easily
check if the requisite Ly$\alpha$\ emission is present. Taking the MOS data
for the full Perseus cluster used in B14, we extended the spectral fit
region from $E=3-6$ keV to $2.55-6$ keV, and included 2 additional
Gaussian components, at $E=2.62$ keV for \ion{S}{15} and at $E=2.96$ keV for
the \ion{Cl}{17} Ly$\alpha$\ line. We do not detect the \ion{Cl}{17}
Ly$\alpha$\ line (which has so far not been observed in any cluster) and can
place a 90\% upper limit on its flux of $5.7\times10^{-6}$ phot cm$^{-2}$
s$^{-1}$. This implies a maximum flux for the Ly$\beta$\ line of $9.5\times
10^{-7}$ phot cm$^{-2}$ s$^{-1}$, conservatively assuming the line ratio of
6.  This is less than 3\% of the maximum allowed flux for the model
\ion{K}{18} line at the same energy in B14 modeling of the same Perseus
spectrum. We note that while fitting the Galactic Center, JP apparently did
not need to include the \ion{Cl}{17} Ly$\alpha$\ line either, in which case
the Cl Ly$\beta$\ should also be negligible. To have a Ly$\beta$\ line in
the absence of the Ly$\alpha$\ line for any ion would be even more exotic
than sterile neutrino.

\section{Conclusions}

We conclude that the JP analysis is severely affected by their use of the
approximate atomic data. When the correct atomic data are used, the line
ratios of S and Ca do not indicate a wide and inconsistent range of
temperatures in clusters, contrary to JP's conclusion. In fact, for the
fiducial and interesting case of Perseus, the S and Ca line ratios --- in
particular, those disentangled from the relative elemental abundances ---
are consistent and indicate a reasonable plasma temperature. They also
exclude significant contributions from cool plasma components to the Ca and
S lines and thus to the possible \ion{K}{18} line (a potential contaminant
for the B14 result) in Perseus and other cluster samples considered in B14.
However, even the \ion{K}{18} fluxes predicted by JP using their atomic data
(excluding the highly implausible estimates based on \ion{Ca}{20}) have
already been allowed by the very conservative B14 fits; the additional
unidentified line was still required. As for the contaminating \ion{Cl}{17}
line proposed by JP that was not included in B14 modeling, this would be a
Ly$\beta$\ line with the absent Ly$\alpha$\ line from the same ion, which is
highly unexpected. We conclude that the 3.5 keV line detection in B14 is not
affected --- with the detailed caveats given in B14.

We have concentrated on the galaxy cluster analysis in JP, although the bulk
of their paper deals with Galactic Center. The incorrect use of atomic data
will affect all of the results; in particular, it may lead to an incorrect
conclusion that a large range of temperatures is not only possible, but
required, to model both the Galactic Center and other systems. As it
happens, the conservative fitting procedure used by B14 makes this
irrelevant for the analysis in B14, but for the benefit of the researchers
studying other objects, we include the correct data in the Appendix.

\section{Appendix: Line Selection}
\label{sec:lines}


We have no reason to doubt the fluxes shown in Table 2 of JP. However, their
choice of lines to include when converting the observed line fluxes into
predicted \ion{K}{18} fluxes for Table 3 is questionable. As an example, for
the He-like \ion{Ar}{17} line they include lines at 3.124, 3.126 and
3.140\,keV, the resonance and two intercombination lines of the system. The
forbidden line at 3.104\,keV, well inside typical CCD energy resolution, was
omitted, although it carries $\approx 25$\% of the flux of this triplet.

Conversely, for the highest energy line in the sample, the Ca XX line at
4.1\,keV, they have included the Ca XX $2p-1s$ doublet, along with the
$1s7p-1s^2, 1s8p-1s^2,1s9p-1s^2$ and $1s10p-1s^2$ transitions of Ar XVII ,
and the \ion{K}{18} $1s3p-1s^2$ transition. These extra lines are largely
included for the low temperature limit, where JP claims the Ar lines
``dominate'', yet the intervening $1s3p-1s^2, 1s4p-1s^2, 1s5p-1s^2$ and
$1s6p-1s^2$ lines at 3.68, 3.87, 3.97 and 4.01\,keV
respectively are not noted as being stronger than the 4.1keV line, which they must be if they originate from Ar.

In Table \ref{tab:jpflux} we show the values of the emissivity of the
\ion{K}{18} $1s2p-1s^2$ triplet predicted by these line fluxes listed in
Table 2 of JP for the MOS detector. The top half shows the emissivities as
calculated using their line list, the bottom using ours, using AtomDB 2.0.2.
The exact lines that we have included, compared with those included by JP,
are listed in Table \ref{tab:linelist}.

\begin{table}[b]
\caption{Revised fluxes (ph cm$^{-2}$ s$^{-1}$) for JP Table 3 of the
  \ion{K}{18} $n=2\rightarrow 1$ triplet based on the Galactic Center
  fluxes presented in JP Table 2, using AtomDB 2.0.2 \label{tab:jpflux}}
\begin{tabular}{cccccc}
Te(keV) & \ion{S}{16} & \ion{Ar}{17} &\ion{Ar}{17} &\ion{Ca}
{19} &\ion{Ca}{20}\\
  &  & (3.13keV) & (3.69keV) & &\\
\hline
\multicolumn{6}{c}{Using original JP lines}\\
0.8 & 6.1e-06 & 8.0e-06 &1.9e-05& 2.4e-05 &1.6e-04\\
1.0 & 3.6e-06 & 9.5e-06 &1.9e-05& 2.0e-05 &1.2e-04\\
2.0 & 2.1e-06 & 1.7e-05 &2.3e-05& 1.5e-05 &2.4e-05\\
5.0 & 1.3e-06 & 3.7e-05 &4.1e-05& 1.2e-05 &1.8e-06\\
\multicolumn{6}{c}{Using our recommended lines}\\
0.8 &6.1e-06 &5.8e-06 &1.8e-05 &2.4e-05 &2.6e-03\\
1.0& 3.5e-06 &6.8e-06 &1.7e-05 &2.0e-05 &1.2e-03\\
2.0 &1.8e-06 &1.1e-05 &1.9e-05 &1.3e-05 &2.6e-05\\
5.0 &7.9e-07 &1.8e-05 &2.4e-05 &7.2e-06 &1.1e-06\\
\hline
\end{tabular}
\end{table}

\begin{table}[t]
\caption{\label{tab:linelist}The lines included in the 6 line complexes identified by JP, in their work and in B14. }
\begin{tabular}{lll}
Feature & Lines (JP) & Lines (This work)\\
\ion{S}{16} & $2p$ $ {}^2P_{1/2}\rightarrow 1s$ $ {}^2S_{1/2}$ & $2p $ ${}^2P_{1/2}\rightarrow 1s $ ${}^2S_{1/2}$ \\
            & $2p$ $ {}^2P_{3/2}\rightarrow 1s$ $ {}^2S_{1/2}$ & $2p $ ${}^2P_{1/2}\rightarrow 1s $ ${}^2S_{1/2}$ \\
\hline
\ion{Ar}{17} \footnote{Ar XVII 3.13 was included in the models of B14 but was not used to constrain the \ion{K}{18} emissivity} & $1s2p$ ${}^1P_{1}\rightarrow 1s^2$ ${}^1S_0$ & $1s2p $ ${}^1P_{1}\rightarrow 1s^2$ $ {}^1S_0$ \\
3.13\,keV            & $1s2p$ ${}^3P_{2}\rightarrow 1s^2$ ${}^1S_0$ & $1s2p  $ ${}^3P_{2}\rightarrow 1s^2$ $ {}^1S_0$ \\
            & $1s2p$ ${}^3P_{1}\rightarrow 1s^2$ ${}^1S_0$ & $1s2p $ ${}^3P_{1}\rightarrow 1s^2$ ${}^1S_0$ \\
            &                                          & $1s2s $ ${}^3S_{1}\rightarrow 1s^2$ $ {}^1S_0$ \\
\hline
\ion{Ar}{17} & $1s3p$ ${}^1P_{1}\rightarrow 1s^2$ ${}^1S_0$ & $1s3p $ ${}^1P_{1}\rightarrow 1s^2$ $ {}^1S_0$ \\
3.69\,keV     &                                                  & $1s3p  $ ${}^3P_{1}\rightarrow 1s^2$ $ {}^1S_0$ \\
\hline
\ion{Ca}{19} & $1s2p$ ${}^1P_{1}\rightarrow 1s^2$ ${}^1S_0$ & $1s2p $ ${}^1P_{1}\rightarrow 1s^2$ $ {}^1S_0$ \\
            & $1s2p$ ${}^3P_{2}\rightarrow 1s^2$ ${}^1S_0$ & $1s2p  $ ${}^3P_{2}\rightarrow 1s^2$ $ {}^1S_0$ \\
            & $1s2p$ ${}^3P_{1}\rightarrow 1s^2$ ${}^1S_0$ & $1s2p $ ${}^3P_{1}\rightarrow 1s^2$ ${}^1S_0$ \\
            &$1s2s $ ${}^3S_{1}\rightarrow 1s^2$ $ {}^1S_0$ & $1s2s $ ${}^3S_{1}\rightarrow 1s^2$ $ {}^1S_0$ \\
\hline
\ion{Ca}{20} & \multicolumn{2}{c}{\ion{Ca}{20}}\\
             & $2p$ $ {}^2P_{1/2}\rightarrow 1s$ $ {}^2S_{1/2}$ & $2p $ ${}^2P_{1/2}\rightarrow 1s $ ${}^2S_{1/2}$ \\
             & $2p$ $ {}^2P_{3/2}\rightarrow 1s$ $ {}^2S_{1/2}$ & $2p $ ${}^2P_{3/2}\rightarrow 1s $ ${}^2S_{1/2}$ \\
             & \multicolumn{2}{c}{\ion{Ar}{17}}\\
             & $1s7p$ $ {}^1P_{1}\rightarrow 1s^2$ $ {}^1S_{0}$ &  \\
             & $1s8p$ $ {}^1P_{1}\rightarrow 1s^2$ $ {}^1S_{0}$ &  \\
             & $1s9p$ $ {}^1P_{1}\rightarrow 1s^2$ $ {}^1S_{0}$ &  \\
             & $1s10p$ $ {}^1P_{1}\rightarrow 1s^2$ $ {}^1S_{0}$ &  \\
             & \multicolumn{2}{c}{\ion{K}{18}}\\
             & $1s3p$ $ {}^1P_{1}\rightarrow 1s^2$ $ {}^1S_{0}$ &  \\
\hline

\ion{K}{18} & \multicolumn{2}{c}{\ion{K}{18}}\\
            & $1s2p$ ${}^1P_{1}\rightarrow 1s^2$ ${}^1S_0$ & $1s2p $ ${}^1P_{1}\rightarrow 1s^2$ $ {}^1S_0$ \\
            &$1s2s $ ${}^3S_{1}\rightarrow 1s^2$ $ {}^1S_0$ & $1s2s $ ${}^3S_{1}\rightarrow 1s^2$ $ {}^1S_0$ \\
            &             & $1s2p  $ ${}^3P_{2}\rightarrow 1s^2$ $ {}^1S_0$
            \footnote{These lines were not included in B14 as the forbidden
              and resonance lines were modeled as separate component, and
              their flux is $\approx$ 10\% of those. However, as JP use a
              single Gaussian for the feature, their flux should be
              included, and they have been in this comment}\\ 
            &                 & $1s2p $ ${}^3P_{1}\rightarrow 1s^2$ ${}^1S_0$ $^{b}$\\
            & \multicolumn{2}{c}{\ion{Cl}{17}}\\
            &  $3p $ ${}^2P_{3/2}\rightarrow 1s$ ${}^2S_{1/2}$ &\\
            &  $3p $ ${}^2P_{1/2}\rightarrow 1s$ ${}^2S_{1/2}$ &\\
            
\end{tabular}
\end{table}

\end{document}